\documentclass[preprint,12pt]{elsarticle}

\usepackage{amssymb}

\journal{Nuclear Inst. and Methods in Physics Research, A}

\usepackage{HGTD_TDR-defs}
\usepackage{multirow} 
\usepackage{rotating}
\usepackage{mathrsfs}
\usepackage{booktabs}
\usepackage{longtable}
\usepackage{placeins}
\usepackage{amssymb}
\usepackage{upgreek}
\usepackage{svg}
\usepackage{comment}
\usepackage{pdfpages}
\usepackage{siunitx}
\usepackage{lineno}
\usepackage{float}
\usepackage{lineno}
\usepackage{tikz}
\usepackage{subcaption}
\title{\boldmath TCAD simulation of AC-LGADs}
 
\author[1]{M.~Nizam}
\author[1]{K.-W.~Shin}
\author[1]{S.M. Mazza}
\author[1]{J. Ott}
\author[1]{A.~Seiden}
\author[1]{B.~Schumm}
\author[1]{Y.~Zhao}

\affiliation[1]{SCIPP, University of California Santa Cruz, 1156 high street, Santa Cruz (CA), US}

\usepackage{sfmath}
\newcommand{\textsp}[1]{$^{\mathrm{#1}}\;$}

\newcommand{\TM}{$^{\mathrm{TM}}\;$}
\newcommand{\Synopsys}{Synopsys$^{\mathrm{TM}}\;$}
\newcommand{\um}{$\mathrm{\mu m}$}

\begin{document}

\begin{frontmatter}

\begin{abstract}
Low Gain Avalanche Detectors (LGADs) are thin silicon detectors with moderate internal signal amplification and time resolution as good as 17 ps for minimum ionizing particles. However, the current major limiting factor in granularity is due to protection structures preventing breakdown caused by high electric fields at the edge of the segmented implants. 
This structure, called Junction Termination Extension (JTE), causes a region of 50-100~\um\ of inactive space. Therefore, the granularity of LGAD sensors is currently limited to the mm scale. 
A solution would be AC-coupled LGADs (AC-LGADs) which can provide spatial resolution on the 10‘s of um scale. This is achieved by an un-segmented (p-type) gain layer and (n-type) N-layer, and a dielectric layer separating the metal readout pads. 
The high spatial precision is achieved by using the information from multiple pads, exploiting the intrinsic charge-sharing capabilities of the AC-LGAD provided by the common N-layer. 
TCAD software (Silvaco and Sentaurus) was used to simulate the behavior of AC-LGADs.
A set of simulated devices was prepared by changing several parameters in the sensor: N+ resistivity, strip length, and bulk thickness.
\end{abstract}


\end{frontmatter}


\section{Introduction}
\label{sec:intro}

Low Gain Avalanche Detectors (LGADs) are thin silicon detectors~\cite{bib:LGAD} capable of providing measurements of minimum-ionizing particles with time resolution as good as 17 ps.
These properties make LGADs the prime candidate technology for achieving 4D tracking in future experiments. 
Furthermore, the fast rise time and short full charge collection time (as low as 1~ns) of LGADs are suitable for high repetition rate measurements in photon science and other fields.
Granularity in traditional DC-LGADs is limited to the mm scale due to protection structures preventing breakdown caused by high electric fields at the edge of the segmented implants. The structure, called Junction Termination Extension (JTE), causes a region of 50-100~\um\ of inactive space between electrodes.

\begin{figure}[ht]
    \centering
    \includegraphics[width=0.55\textwidth]{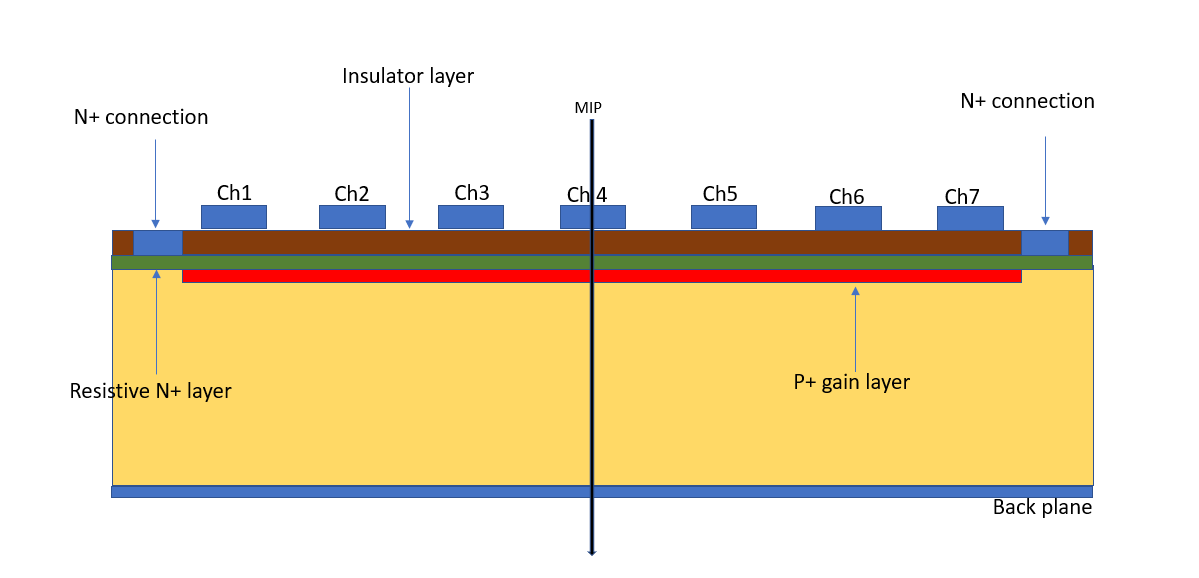}
    \includegraphics[width=0.44\textwidth]{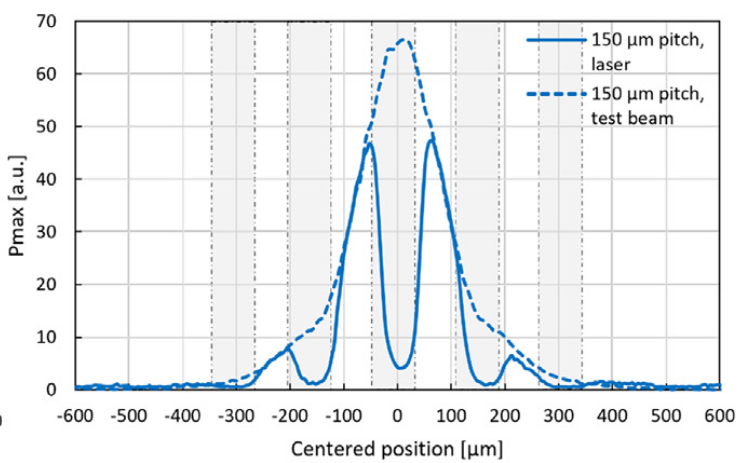}
    \caption{Left: AC-coupled Low Gain Avalanche Diode (AC-LGAD) schematic. Right: charge sharing profile around one strip measured with laser TCT (electrode metal is opaque, so the response under the strip cannot be probed) and test beam of a strip AC-LGAD with a pitch of 200~\um\ and strip width of 80~\um. Strips are indicated by the grey overlay. Plot is from reference~\cite{OTT2023167541}.}
    \label{fig:aclgaddraw}
\end{figure}

AC-coupled LGADs~\cite{OTT2023167541} (AC-LGADs, also named Resistive Silicon Detectors, RSD) overcome the granularity limitation of traditional LGADs and have been shown to provide a spatial resolution of the order of 10s of~\um\. 
This remarkable feature is achieved with an un-segmented (p-type) gain layer and a resistive (n-type) N-layer. 
An insulating dielectric layer separates the metal readout pads from the N+ resistive layer as shown in Fig.~\ref{fig:aclgaddraw}, Left. 
The high spatial precision is achieved by using the information from multiple metal pads, exploiting the intrinsic charge-sharing capabilities of the AC-LGAD provided by the common resistive N-layer. 
AC-LGADs are the chosen technology for near-future large-scale applications like the EPIC detector at the Electron-Ion Collider at BNL and the PIONEER~\cite{Mazza:2021adt} experiment.

\label{sec:silvaco}

Charge-sharing in AC-LGADs depends on a series of parameters: resistivity of the N+ layer, geometry and pitch of the metal electrodes, bulk thickness, and oxide layer thickness.
This paper will present a set of TCAD simulations on AC-LGAD devices. 
Two TCAD simulation software, TCAD Sentaurus, and TCAD Silvaco, were used to understand the charge-sharing behavior of AC-LGADs as a function of the detector parameters.
TCAD Silvaco was used in 2D approximation, while TCAD Sentaurus was used with a full 3D simulation.
The simulated sensors are strip AC-LGADs with a bulk thickness of 50~\um\ and 120~\um\, strip pitch of 200~\um\, and the width of each strip is 80~\um\.
Test beam data from a prototype sensor with the same geometry fabricated at BNL was used to tune the simulation parameters.
This device's measured charge sharing profile is shown in Fig.~\ref{fig:aclgaddraw}, Right.
In the simulated results, the charge-sharing profile is only simulated from the center of the strip toward one direction.
The following sections will explain the simulation setup and show the resulting simulated charge-sharing profiles.

\section{2D Simulation of AC-LGAD}
\begin{figure}[ht]
    \centering
    \subfloat[\centering]{{\includegraphics[height=5.0cm, width=6.5cm]{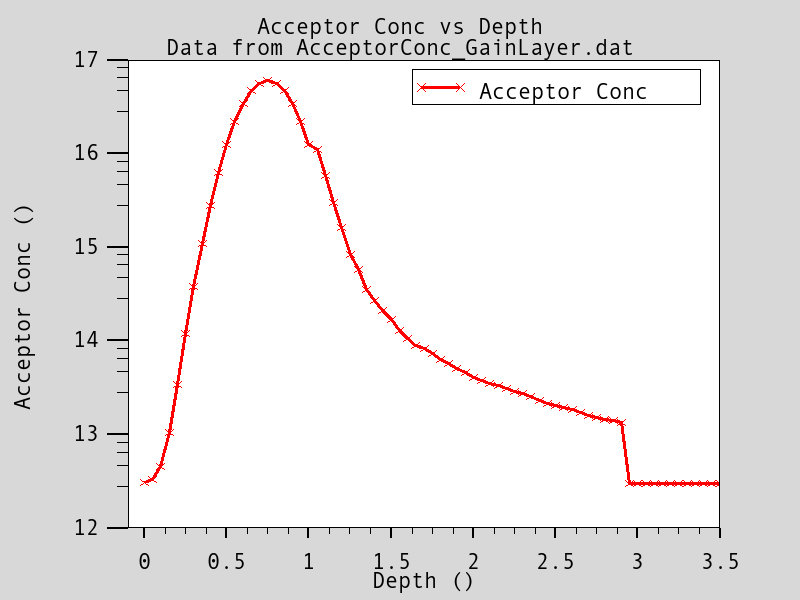} }}%
    \quad
    \subfloat[\centering]{{\includegraphics[height=5.0cm, width=6.5cm]{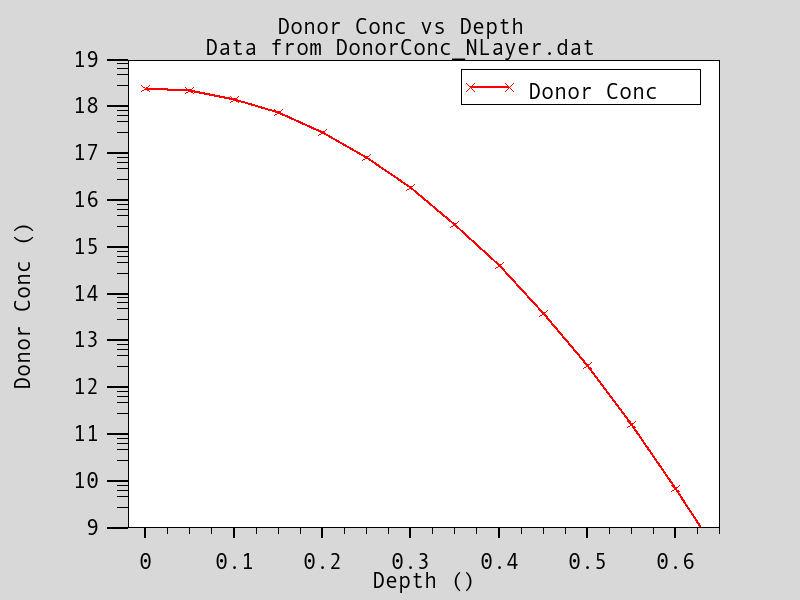} }}%
    \caption{(a) Acceptor impurity concentration in the gain layer as a function of depth. (b) Donor impurity concentration in the N+ layer as a function of depth.}%
    \label{fig:TCAD_pro}%
\end{figure}

We studied the effects of resistivity of the N+ layer and pitch size on signal sharing between neighboring channels using 2D Silvaco$^{\copyright}$ TCAD tools. A 50$\mu$m thick and seven-channel AC-LGAD device shown in Fig.~\ref{fig:aclgaddraw} (left) was simulated for different pitch sizes and N+ resistivity. The doping profile of the gain layer was kept the same for all the cases and an analytical model was used for the N+ layer profile to achieve different N+ resistivities. The doping profiles extracted from the simulated device are shown in Fig.~\ref{fig:TCAD_pro}. The p-type doping concentration in the gain layer as a function of depth is shown Fig.~\ref{fig:TCAD_pro}(a) and the n-type doping concentration in N+ layer is shown in Fig.~\ref{fig:TCAD_pro}(b). The doping concentration in the p-type bulk was 2 $\times$10$^{13}$/cm$^{3}$ and the peak doping concentration in the gain layer was 6$\times$10$^{16}$/cm$^{3}$.
The resistivity of N+ layer is tuned by varying peak doping concentrations. IV characteristics for different N+ doping concentrations is shown in figure \ref{fig:aclgadIV}. Single Event Upset method was used to generate 80 electron-hole pairs per micron to simulate Minimum Ionising Particle (MIP) response from the device. 
An impact ionization model called "GRANT" available in Silvaco victorydevice is used to generate electron-hole pairs. It was developed after the investigation of Baraff~\cite{Nizam:BARAFF} which incorporates low, intermediate and high field response regions for the electron and hole ionization rates. The coefficient values are chosen to match the experimental data of Grant \cite{Nizam:GRANT19731189}. A doping concentration dependent mobility model "CONMOB" is used along with "FLDMOB" which takes care of the saturation drift velocities of electrons and holes in silicon. The SRH recombination model derived by Shockley and Read \cite{nizam:SR} and Hall \cite{nizam:hall} is used which depends on carrier lifetimes.
\begin{figure}[ht]
    \centering
    \includegraphics[width=0.875\textwidth]{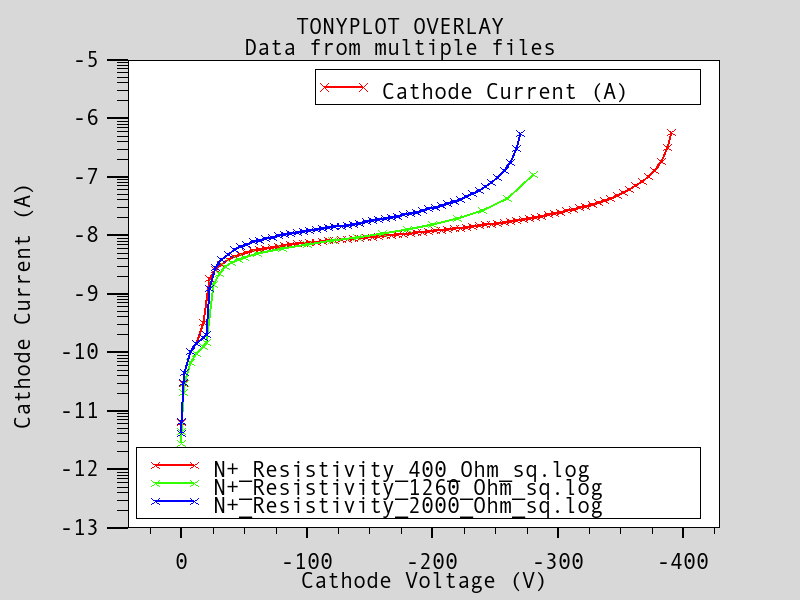}
    \caption{IV characteristics of simulated devices for different N+ layer doping concentrations.}
    \label{fig:aclgadIV}
\end{figure}
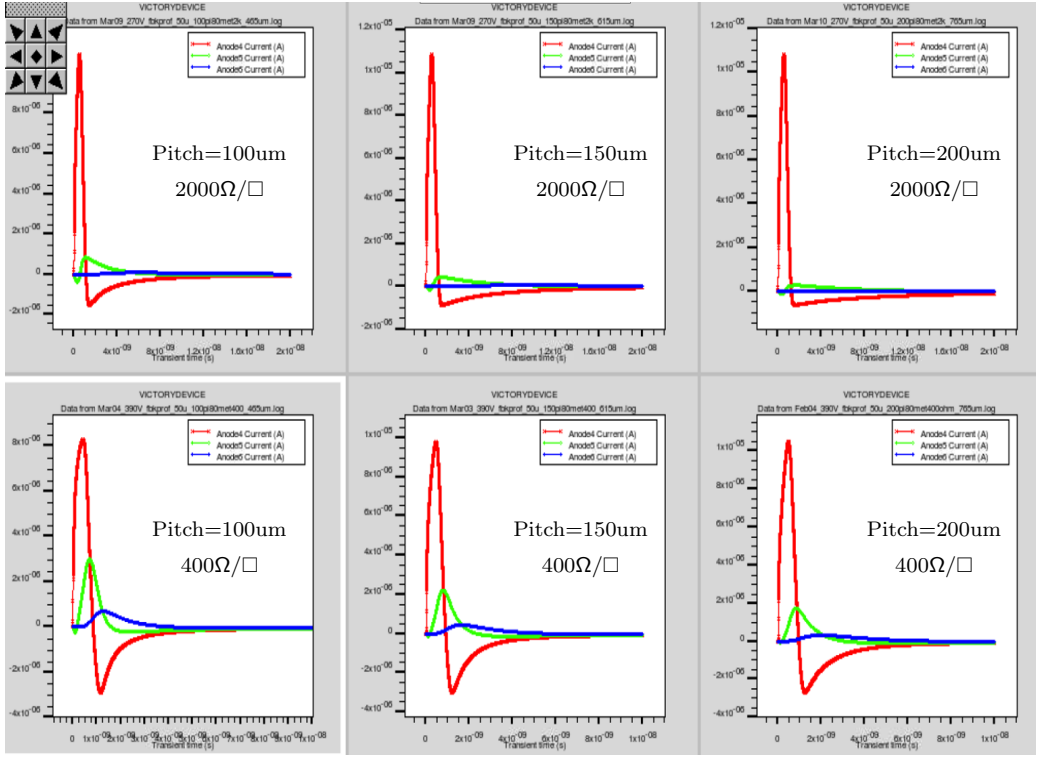
\begin{figure}[ht]
\begin{tikzpicture}
    \draw (0, 0) node[inner sep=0] {\includegraphics[height=10.0cm, width=13.7cm]{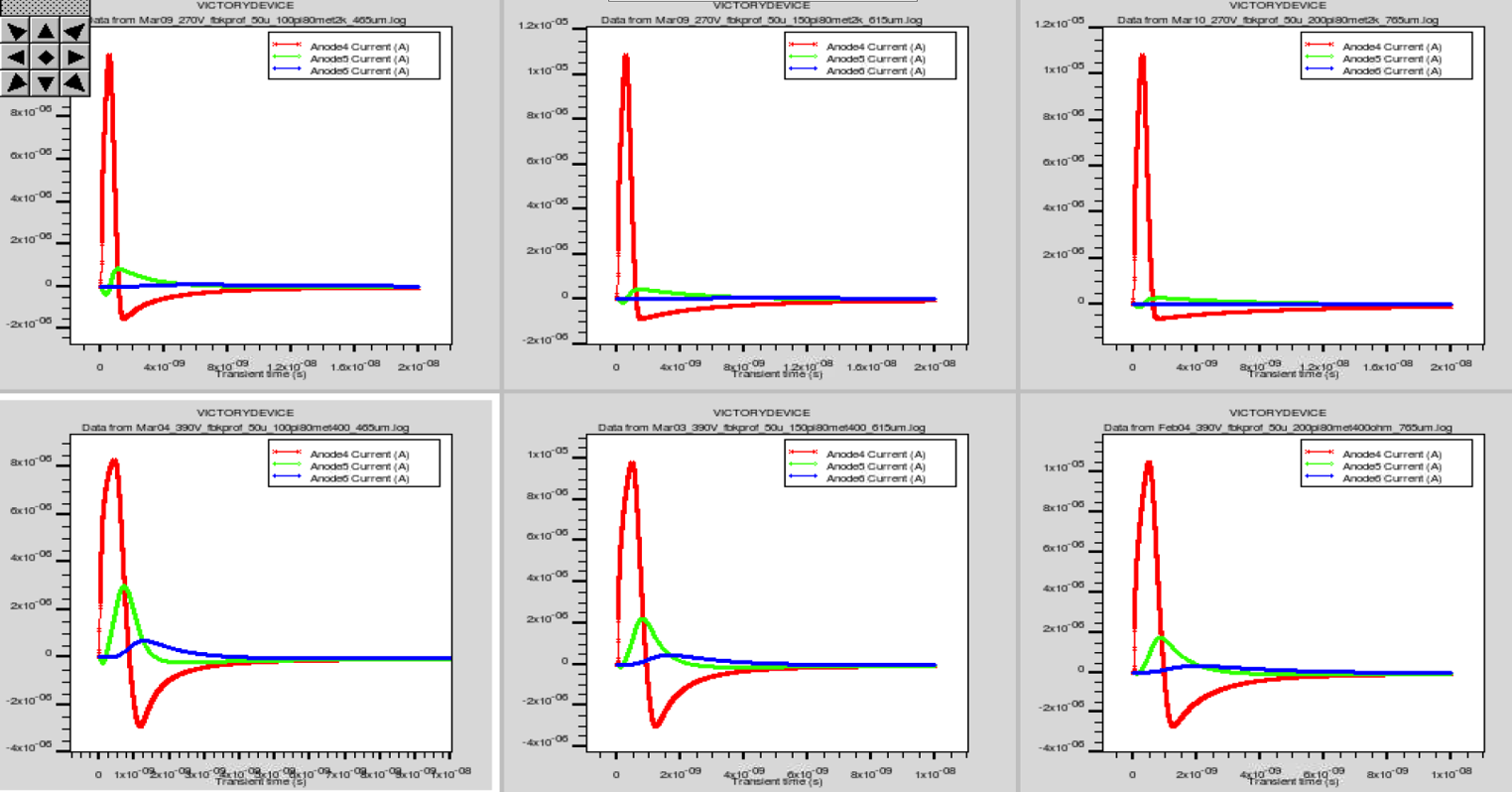}};
    \draw (-4, 3) node {\scriptsize Pitch=100um};
    \draw (-4, 2.5) node {\scriptsize 2000$\Omega/\square$};
    \draw (-4, -2) node {\scriptsize Pitch=100um};
    \draw (-4, -2.5) node {\scriptsize 400$\Omega/\square$};
    \draw (0.8, 3) node {\scriptsize Pitch=150um};
    \draw (0.8, 2.5) node {\scriptsize 2000$\Omega/\square$};
    \draw (0.8, -2) node {\scriptsize Pitch=150um};
    \draw (0.8, -2.5) node {\scriptsize 400$\Omega/\square$};
    \draw (5.5, 3) node {\scriptsize Pitch=200um};
    \draw (5.5, 2.5) node {\scriptsize 2000$\Omega/\square$};
    \draw (5.5, -2) node {\scriptsize Pitch=200um};
    \draw (5.5, -2.5) node {\scriptsize 400$\Omega/\square$};
\end{tikzpicture}
\caption{Simulated signal waveforms for pitch sizes 100$\mu$m, 150$\mu$m and 200$\mu$m for 400$\Omega/\square$ and 2000$\Omega/\square$ N+ resistivity. Red is main channel, green is first neighbour and blue is second neighbour.}
\label{fig:siwave}
\end{figure}

The simulated device has bulk thickness of 50~$\mu$m and seven channels with each electrode of 80$\mu$m width. We simulated three pitch sizes 100~$\mu$m, 150~$\mu$m, and 200~$\mu$m to compare with the test beam data of a similar sensor developed at the Brookhaven National Laboratory (BNL) and tested at the Fermilab test beam and with laser TCT (Fig.~\ref{fig:aclgaddraw}, Right). We also performed a systematic study of N+ resistivity and its effects on signal sharing. In a seven-channel device, MIP was injected normally at the center of the middle channel (4$^{th}$ channel). Signals at the main channel, first neighbor, and second neighbor for the three pitch sizes and the two sample N+ resistivity 400~$\Omega/\square$ and 2000~$\Omega/\square$ are shown in Fig.~\ref{fig:siwave}. Anode4 (red) is the main channel and Anode5 (green) and Anode6 (blue) are the first and second neighbors. Pulse amplitude for Anode5 and Anode6 as the pitch size is increased from 100~$\mu$m to 200~$\mu$m. The neighboring channel's pulse amplitude decreases when N+ resistivity is increased from 400 to 2000 $\Omega/\square$.

\begin{figure}[ht]
    \centering
    \subfloat[\centering]{{\includegraphics[height=5.0cm, width=6.5cm]{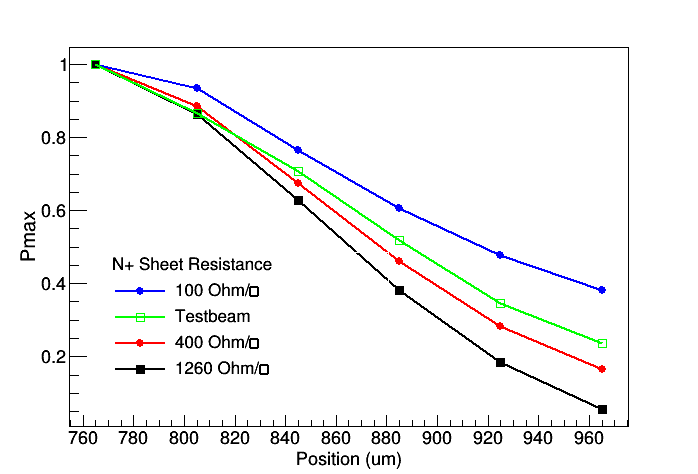} }}%
    \quad
    \subfloat[\centering]{{\includegraphics[height=5.0cm, width=6.5cm]{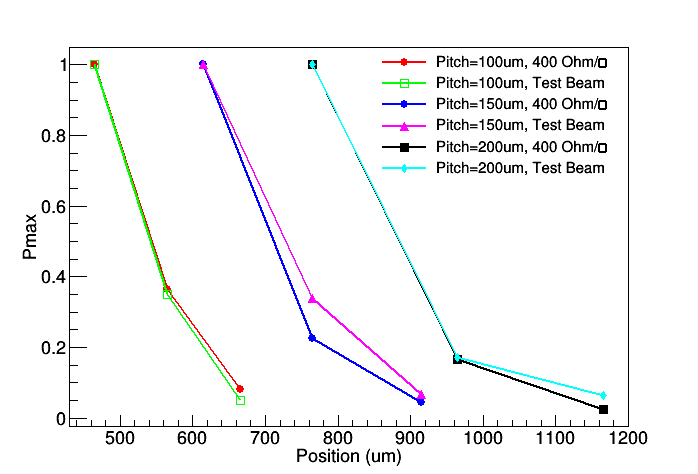} }}%
    \caption{(a) Fraction of Pmax for different N+ resistivity. (b) Fraction of Pmax for different pitch sizes.}%
    \label{fig:Pmaxfrac}%
\end{figure}

Fractional Pmax values of the signal are obtained as a function of the distance from the location of charge injection and compared with the test beam data of the corresponding BNL sensor geometry. Fig.~\ref{fig:Pmaxfrac}(a) shows fractional Pmax values as a function of the location of the charge injection for different values of the simulated N+ resistivity. It is shown that charge sharing between neighboring channels can be minimized by increasing the resistivity of the N+ layer. We compared our simulated data with the test beam results and found that the 300 - 400~$\Omega/\square$ resistivity of the 2D simulation produces charge 
sharing similar to the experimental data. The charge shared by the first neighbor which is centered at 960 $\mu$m is about 20\% of the main channel. Charge sharing is reduced to below 10\% for resistivity greater than 1000~$\Omega/\square$. We fixed the resistivity at 400~$\Omega/\square$ to study the effect of pitch size on charge sharing. Fig.~\ref{fig:Pmaxfrac}(b) shows fractional Pmax values as a function of the location of the charge injection for different pitch sizes and the corresponding values from test beam data analysis. The pitch size effect on charge sharing can be seen as we go from a 100~$\mu$m pitch to a 200~$\mu$m pitch. Charge sharing in the first neighbor is more than 30\% for a 100~$\mu$m pitch but only $\sim$20\% for a 200~$\mu$m pitch.

\section{3D TCAD Simulation Set Up}
\begin{figure}[ht]
    \centering
    \subfloat[\centering]{{\includegraphics[height=4.0cm, width=7.0cm]{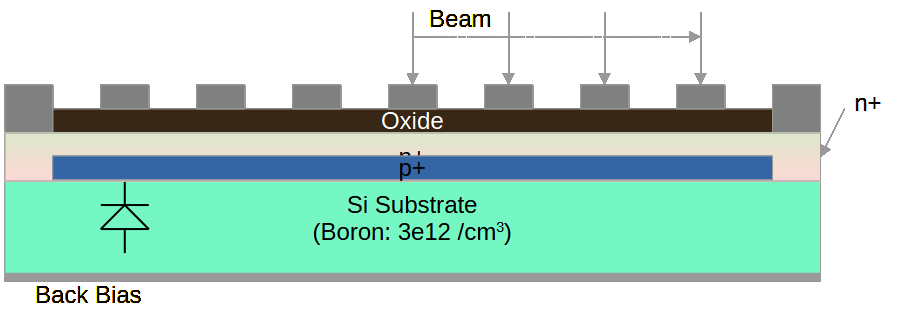} }}%
    \quad
    \subfloat[\centering]{{\includegraphics[height=4.6cm, width=6.0cm]{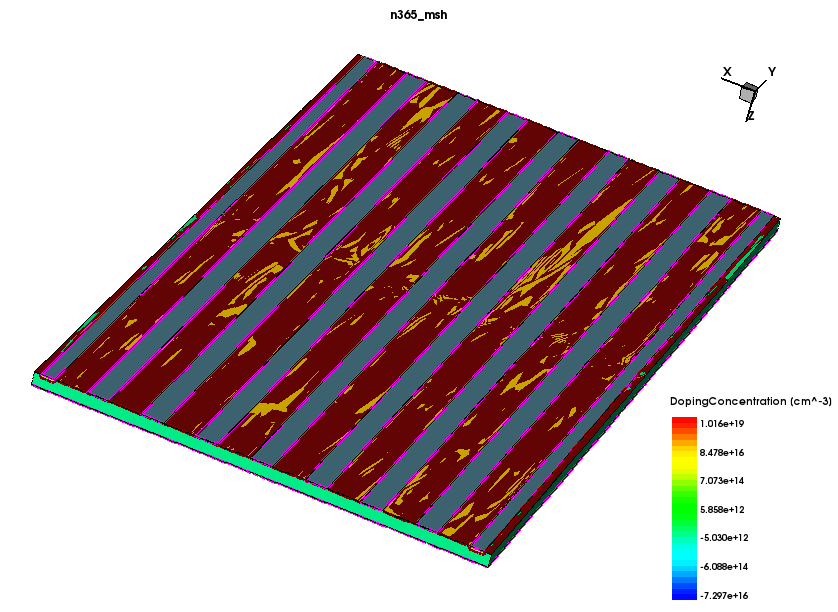} }}%
    \caption{TCAD model of 7 channel devices.(a) Side view. (b) 3D rendering view.}%
    \label{fig:TCAD_Models}%
\end{figure}

To investigate granularity of LGAD devices, a TCAD approach was taken with 3D simulation method. Among a few candidates in the market, we employed \Synopsys version K2015.06-SP2 EDA package. The device simulation was performed by \Synopsys Sentaurus and the TCAD device model was analytically constructed by \Synopsys Device editor. Figure \ref{fig:TCAD_Models} shows the sketch of LGAD strip device and 3D rendering of simulation device model of the AC-LGAD strip detector. 

The structure of the AC-LGAD model is based on a 50-\um-thick silicon chunk with baseline boron doping concentration of 3 $\times$ 10\textsp{12} cm\textsp{-3} to emulate high-resistivity p-type silicon, the resistivity is around 5000 $\Omega \cdot \textrm{cm}$. At the top side of the silicon substrate, additional boron and phosphorus doping profiles from SIMS data, equivalent to 1.2 k$\Omega$/sqare of sheet resistance, provided by Brookhaven National Laboratory, were imported onto the model to describe a typical AC-LGAD with gain layer. The LGAD area was also terminated laterally by extended junction n+ doping profiles which will be making direct contact with the readout electrodes at the right and left ends of the device. 

On top of the doped LGAD layer, the silicon substrate was passivated by 100-nm-thick silicon dioxide material which will be followed by aluminum on top of it. The metal layers are 500-nm-thick aluminum, deposited on top of a silicon dioxide layer or heavily doped silicon surface to depict electrode structure for strip AC-LGAD detector. In the middle, there are seven 80-\um-wide strip electrodes are placed with a pitch of 200-\um. The readout electrodes are 50 um wide each, making direct contact to highly doped n+ at each end of the silicon substrate, effectively making cathode contacts. The highly doped contact n+ profiles are analytically described with peak location at the surface (peak depth = 0 \um) with a concentration of 10\textsp{18}cm\textsp{-3} with Gaussian tail width (standard deviation) of 50 nm. However, to emulate deep n+ JTE area, this contact area was again doped with a phosphorus profile of 10\textsp{16}cm\textsp{-3} of peak concentration, 0 \um\; peak depth, and 1.0 \um\; of Gaussian tail width to enclose the LGAD area with boron implanted gain layer. The other side of the silicon substrate was simply covered with a 0.5-\um-thick aluminum layer to provide a whole device-size anode electrode. The back side doping of the silicon substrate was not necessarily doped since the simulator does not consider a metal contact as Schottky unless designated so. Thus, all the direct silicon contacts were considered ohmic contacts.

\begin{figure}[ht]
    \centering
    \includegraphics[width=0.875\textwidth]{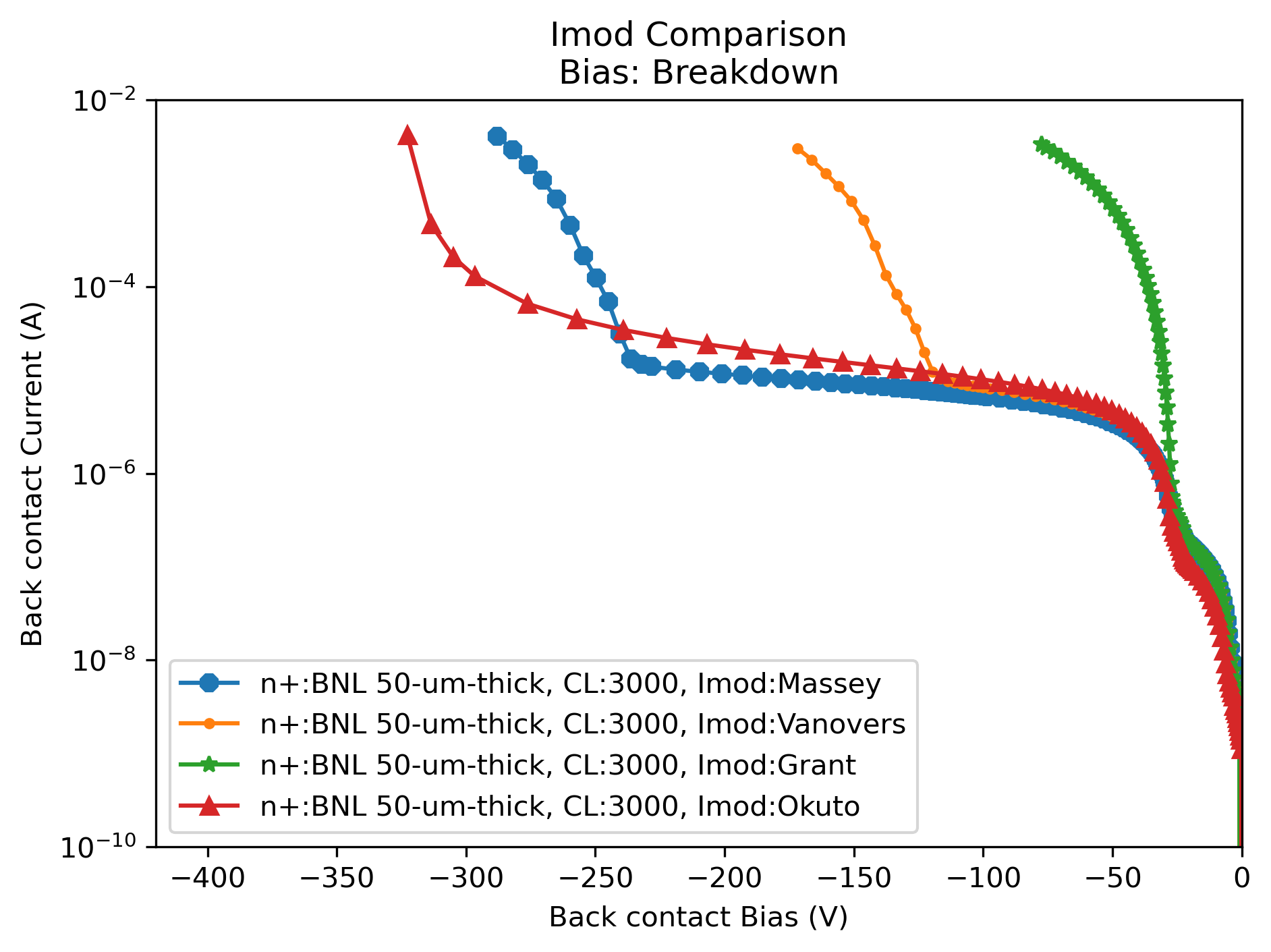}
    \caption{IV sweep results with various impact ionization models.}
    \label{fig:TCAD_LGAD_IVSweep}
\end{figure}

The simulation procedure consists of two phases: voltage biasing and transient response. The voltage biasing phase is a mandatory phase to provide baseline device status of the simulated AC-LGADs before illuminating the detector with a single beam for transient simulation. On the other hand, the biasing simulation provided breakdown point of the simulated AC-LGAD devices which is depicted in Fig. \ref{fig:TCAD_LGAD_IVSweep} with a few different impact ionization models. 

In the simulation, other than the impact ionization models, the electron transport models for high field velocity saturation with doping dependence under 298 K of ambient temperature as well as Schockly-Read-Hall(SRH) recombination model, tuned with doping and temperature dependence. Additionally, the SRH model was backed up with Auger and Band-to-Band recombination models to depict a high-field reverse-biased diode effectively. 

Here, at the impact ionization comparison for 3D AC-LGAD model with 3 mm strip length, Fig. \ref{fig:TCAD_LGAD_IVSweep}, the Okuto-Crowell model (Okuto) appeared to be most relevant to most of the experimental data in terms of breakdown measurement. Another close candidate seems to be Massey model \cite{bib:MarcoRD50June2017} showing a bit earlier breakdown of -250 V of back contact bias. However, Okuto-Crowell model, which was also employed in related previous work \cite{kyungwook2021lgad}, was selected to take advantage of simulation speed since it was built into the Sentaurus while Massey model was implemented via PMI method. On the other hand, another built-in model, van Overstraeten-de Man's model was not selected due to unrealistically low breakdown voltage. The Grant model was working relatively well with Silvaco\TM VictoryDevice simulators in 2D simulation. But in PMI implemented \Synopsys Sentaurus was not the case in 3D model simulation as breakdown at even lower than -50 V of back contact bias which is unrealistic.

The beam was implemented as slit-thin ionization profile (Gaussian radius of 0.1 \um) which was provided with 'HeavyIon' excitation capability of the Sentaurus simulator. The ionization density was chosen as 1.28$\times$10\textsp{-5}pC/\um\; across the silicon substrate to emulate an infrared laser beam. The beam was impinging into the AC-LGAD from directly above from the channel electrodes while switching positions from the middle channel (Channel 4) to channel 7 (the rightmost electrode) as depicted in Fig.~\ref{fig:TCAD_Models} (a). The beam illuminated the AC-LGAD device model for 0.1 ps at the beginning of the transient simulation. 

The numerical convergence issue was amended with a custom ILS method while transient simulation completed with default Backward Euler method without any convergence issues with 10\textsp{-24} second of minimum timing resolution until 10 ns of total transient simulation timing range. The customized ILS setting can be found below.

\begin{verbatim}
ILSrc="set (5) {
  iterative(
    gmres(100), tolrel=1e-8, tolunprec=1e-4,
      tolabs=0, maxit=250);
    preconditioning(ilut(1e-7,-1), left);
    ordering(symmetric=nd, nonsymmetric=mpsilst);
    options(compact=yes, linscale=0, refineresidual=24,
      refineiterate=1, refinebasis=1, verbose=0); };"
\end{verbatim}

In the end, the transient simulation starting point was set up -250~-385~V of back contact bias under Okuto-Crowell avalanche generation model to ensure LGAD gain generation action is present during the transient response simulation of the AC-LGAD.

\section{3D TCAD Simulation Results}

\begin{figure}[ht]
    \centering
    \subfloat[L = 500 \um]{
        \includegraphics[width=0.45\textwidth]{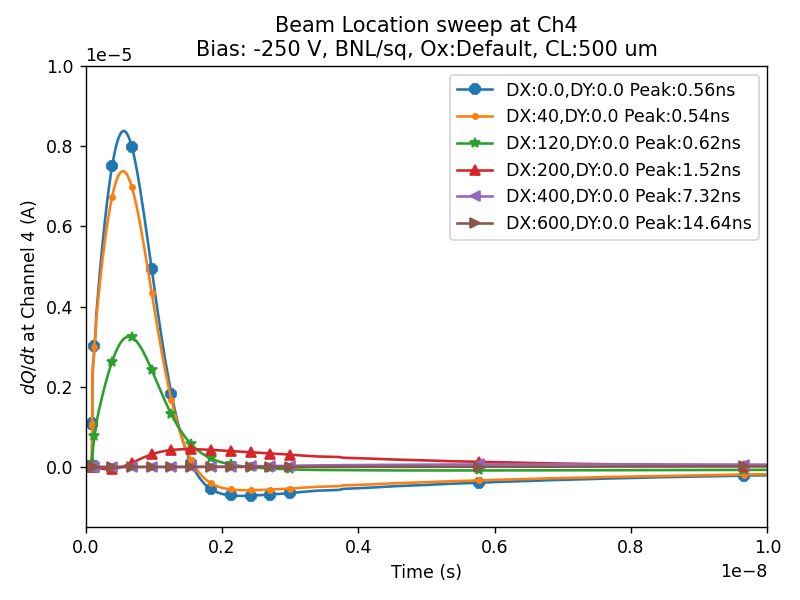}\label{fig:TCAD_Transients_L500}}
    \hfill
    \subfloat[L = 3000 \um]{
        \includegraphics[width=0.45\textwidth]{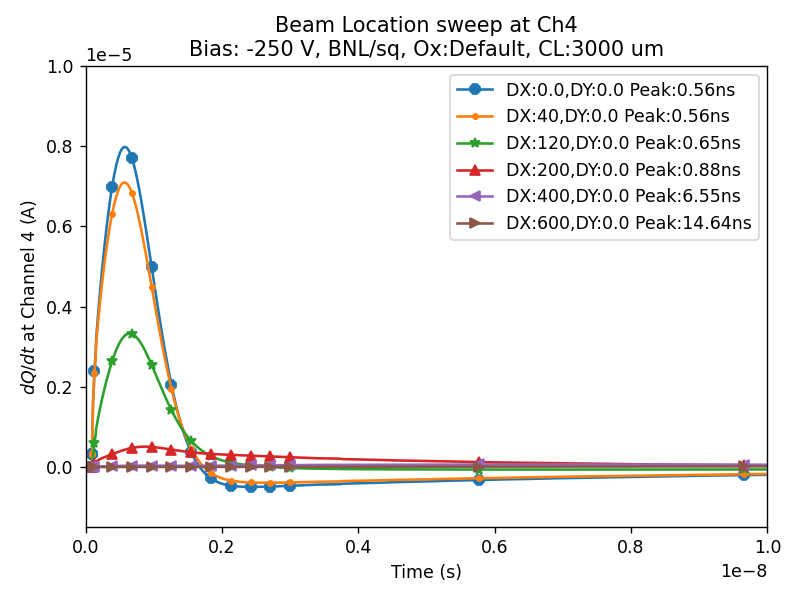}\label{fig:TCAD_Transients_L3000}}
    \vskip\baselineskip
    \subfloat[L = 5000 \um]{
        \includegraphics[width=0.45\textwidth]{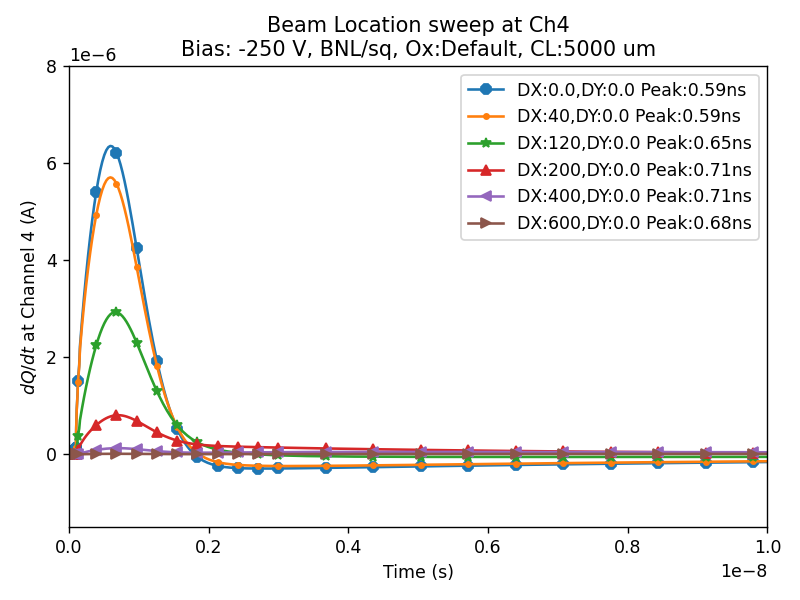}\label{fig:TCAD_Transients_L5000}}
    \hfill
    \subfloat[L = 2 cm]{
        \includegraphics[width=0.45\textwidth]{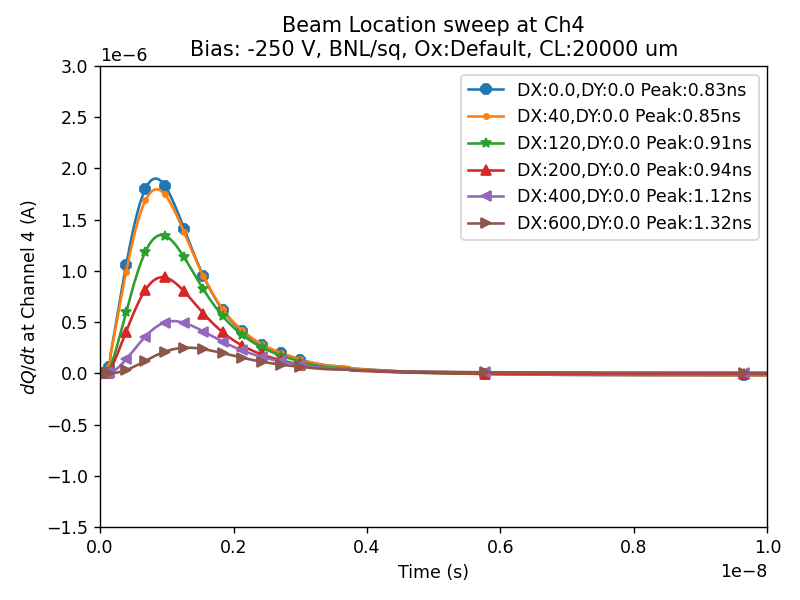}\label{fig:TCAD_Transients_L2cm}}
   
    \caption{Transient responses of varying channel lengths of 50-\um-thick devices.}
    \label{fig:TCAD_Transients}
\end{figure}

\paragraph{Strip Length}
Fig. \ref{fig:TCAD_Transients} summarizes the transient responses from the 7-Channel AC-LGAD device. The granularity is basically extracted from those transient responses. The strip channel length investigation involved 4 different strip length settings: 500~\um, 3000~\um, 5000~\um, and 20000~\um (or 2 cm) devices. As the channel length grows from 500~\um\ to 2 cm, we can observe the peak current intensity gradually decreases due to lengthened collection electrodes. 
The blue curves at each sub-figure in Fig. \ref{fig:TCAD_Transients} indicate the current induced by the MIP, hitting the middle channel (channel 4). 
The induced current overshoots a little bit and converges down to zero Amperes. The end value of the $dQ/dt$ is certainly zero since those current values are not exactly extracted from TCAD results. 
In fact, the TCAD simulator does not actually calculates the current flowing through the AC-LGAD channels since they are literally floating electrodes. Thus, the current values were obtained by taking 1st derivative of induced charges, instead. Hence the notation says $dQ/dt$. The current zero overshoot, negative current at a few cases in Fig. \ref{fig:TCAD_Transients} can also be observed when the beam is hitting 40 \um\ shifted position from the center of channel 4, which is actually hitting the right side edge of the channel 4 electrode.

On the other hand, when the beam is hitting in the middle of two neighboring channels, 120 \um\ shifted beam, green graph in each subplot in Fig.~\ref{fig:TCAD_Transients}, to 3rd neighbors, the inversion disappears except in 2-cm-long channel where all the induced currents are an order of magnitude less than shorter strip devices. Generally, the induced currents at the strips gradually decrease as we increase the channel length. The drop of maximum current is prevalent as we exceed the channel length of 5000 \um.

\begin{figure}[ht]
\centering
\includegraphics[width=0.85\textwidth]{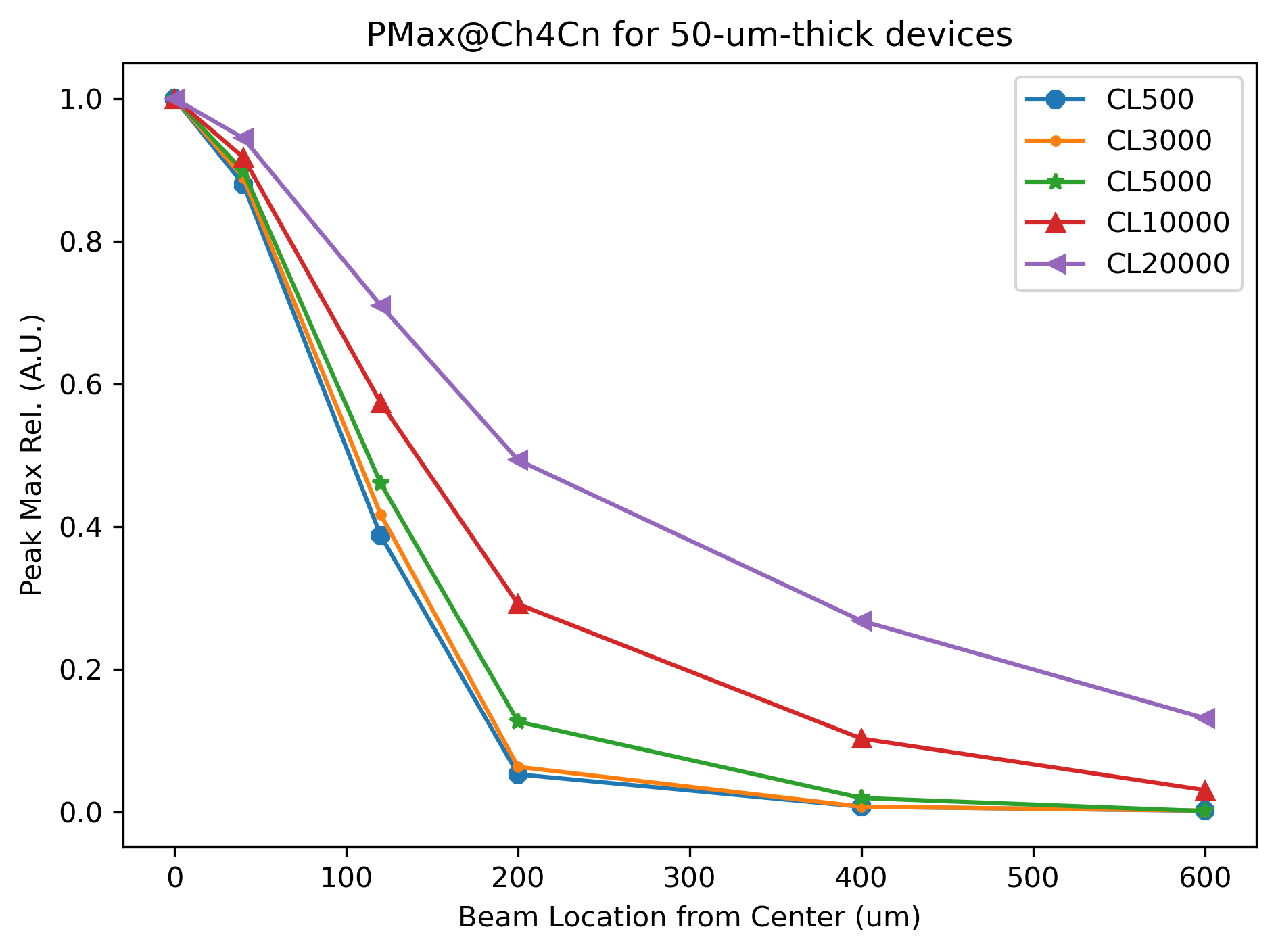}
\caption{P-Max plot of various strip lengths.}
\label{fig:TCAD_PMAX}
\end{figure}

To visualize the granularity more efficiently, Fig. \ref{fig:TCAD_PMAX} was implemented to show the drop of peak currents when the beam is hitting different locations. Here, the lateral axis is the offset from the middle of the channel 4 electrode. Thus, 200 \um, 400 \um, and 600 \um\ locations depict 1st, 2nd, and 3rd neighbors (channels 5, 6, and 7,) respectively. The peak heights are normalized at the original location (the middle of channel 4 electrode) among all the different channel length devices for simpler comparison. 

The 1st neighbor shows from 40 to 50\% of peak height compared to the original location, except the 2-cm-long device where the total device width (1670 \um, perpendicular to strip longitudinal direction) becomes less than 1/10 of the device length, 20000 \um. The 2-cm-long device shows a first neighbor signal ratio of almost 50\% of the original location, almost a quarter of charge sharing between channels than the shorter devices. If we define aspect ratio with the width-to-length ratio, 3.54 for 500-\um-long, 0.56 for 3000-\um-long, 0.334 for 5000-\um-long, and 0.0835 for 2-cm-long devices. In other words, there is a reversely proportional relation between the AC-LGAD charge sharing and the strip channel length. 

To make matters worse in the long strip devices, the 2nd and 3rd neighbors still show 35\% and 20\% of peak height against the original location while the shorter devices show almost no crosstalk from the center device. In short, we can assume that almost no influence is present from 2nd neighbor or further channels when the channel length is not exceeding a certain point. The acceptable threshold of P-Max deviation for the 1st neighbor is under discussion but it is preferable to be as low as less than 20\% of the original position.

\begin{figure}[ht]
    \centering
    \subfloat[L = 500 \um]{
        \includegraphics[width=0.45\textwidth]{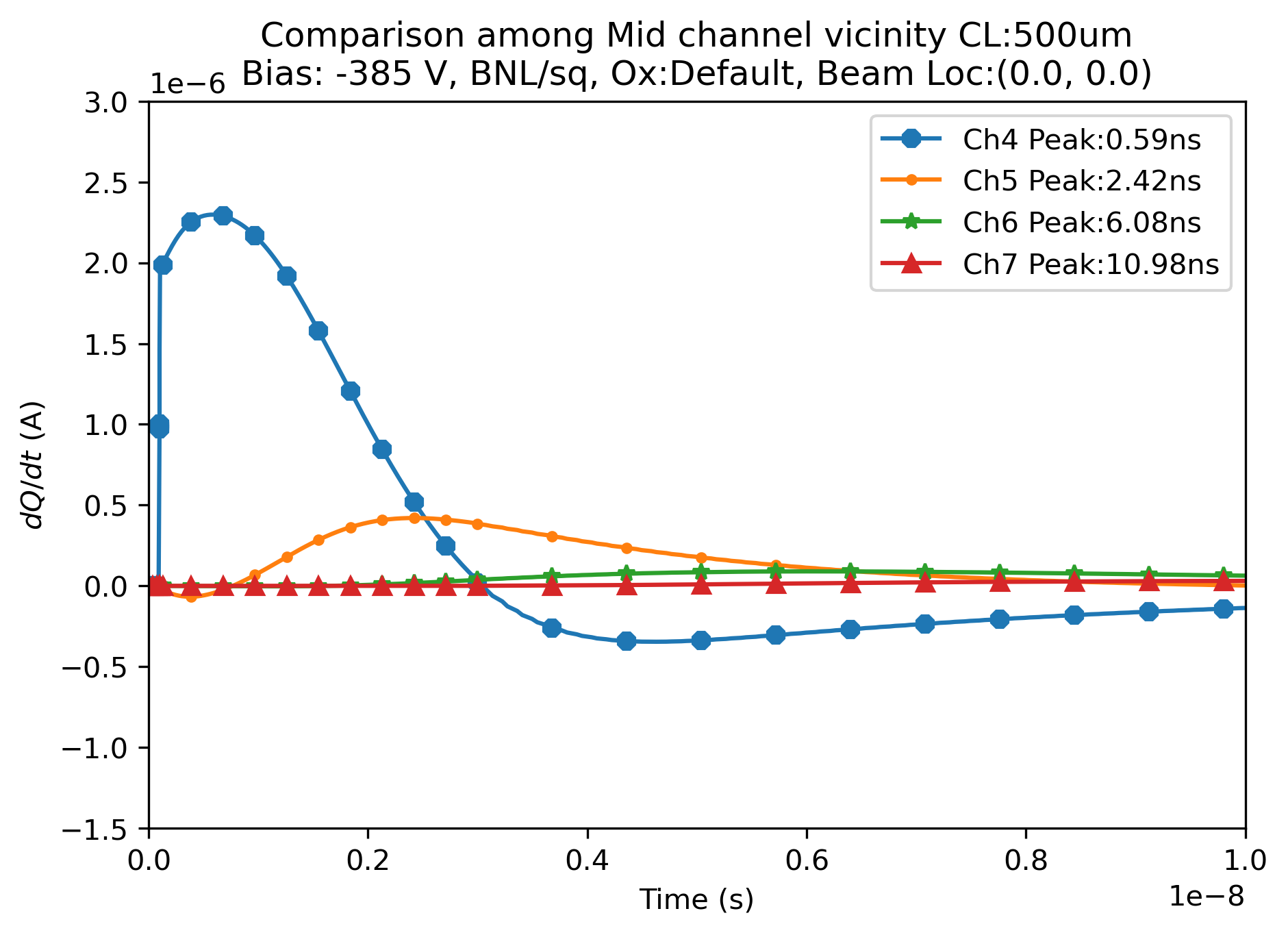}\label{fig:TCAD_Transients_120um_L500}}
    \hfill
    \subfloat[L = 3000 \um]{
        \includegraphics[width=0.45\textwidth]{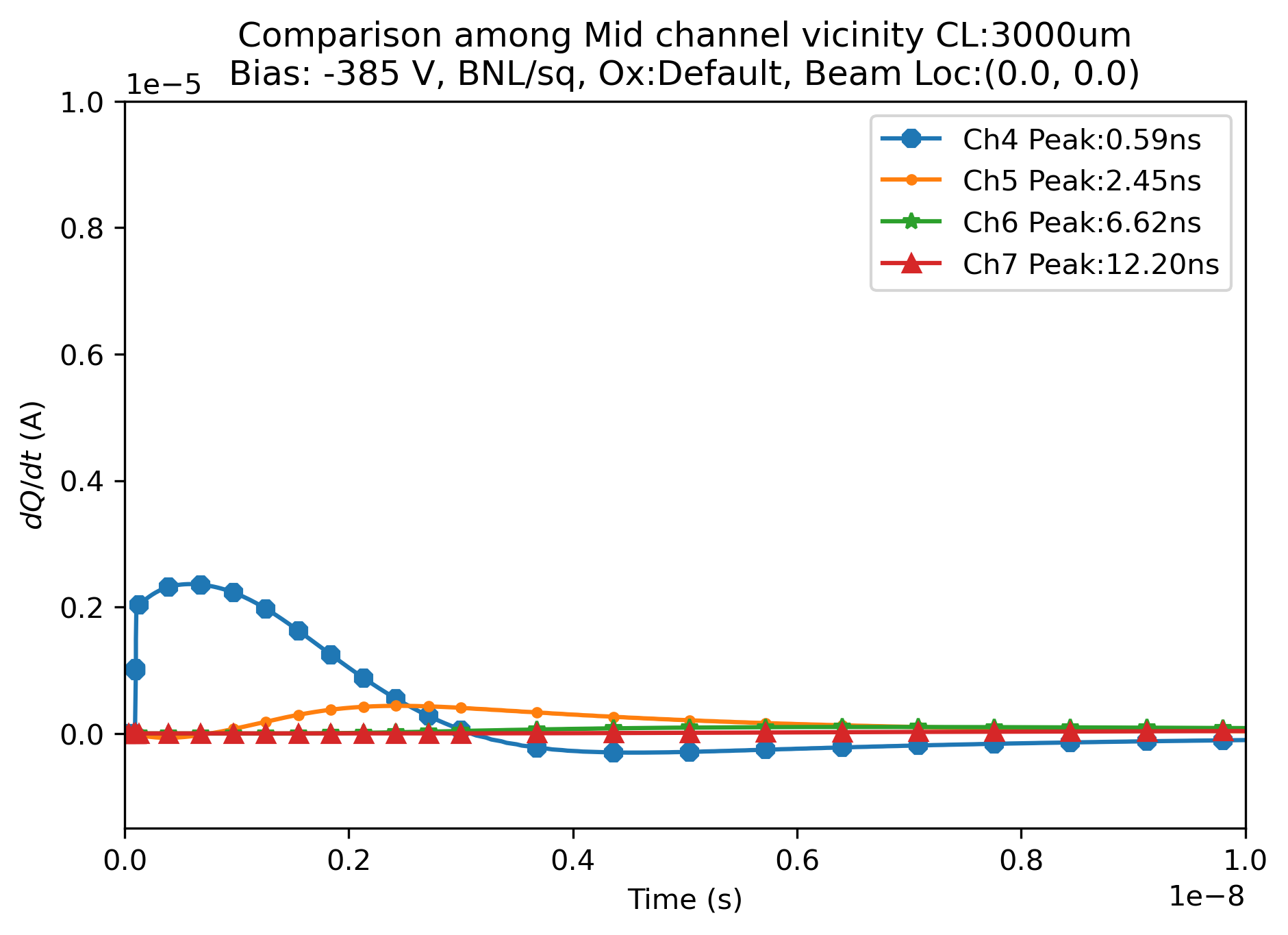}\label{fig:TCAD_Transients_120um_L3000}}
    \vskip\baselineskip
    \subfloat[L = 5000 \um]{
        \includegraphics[width=0.45\textwidth]{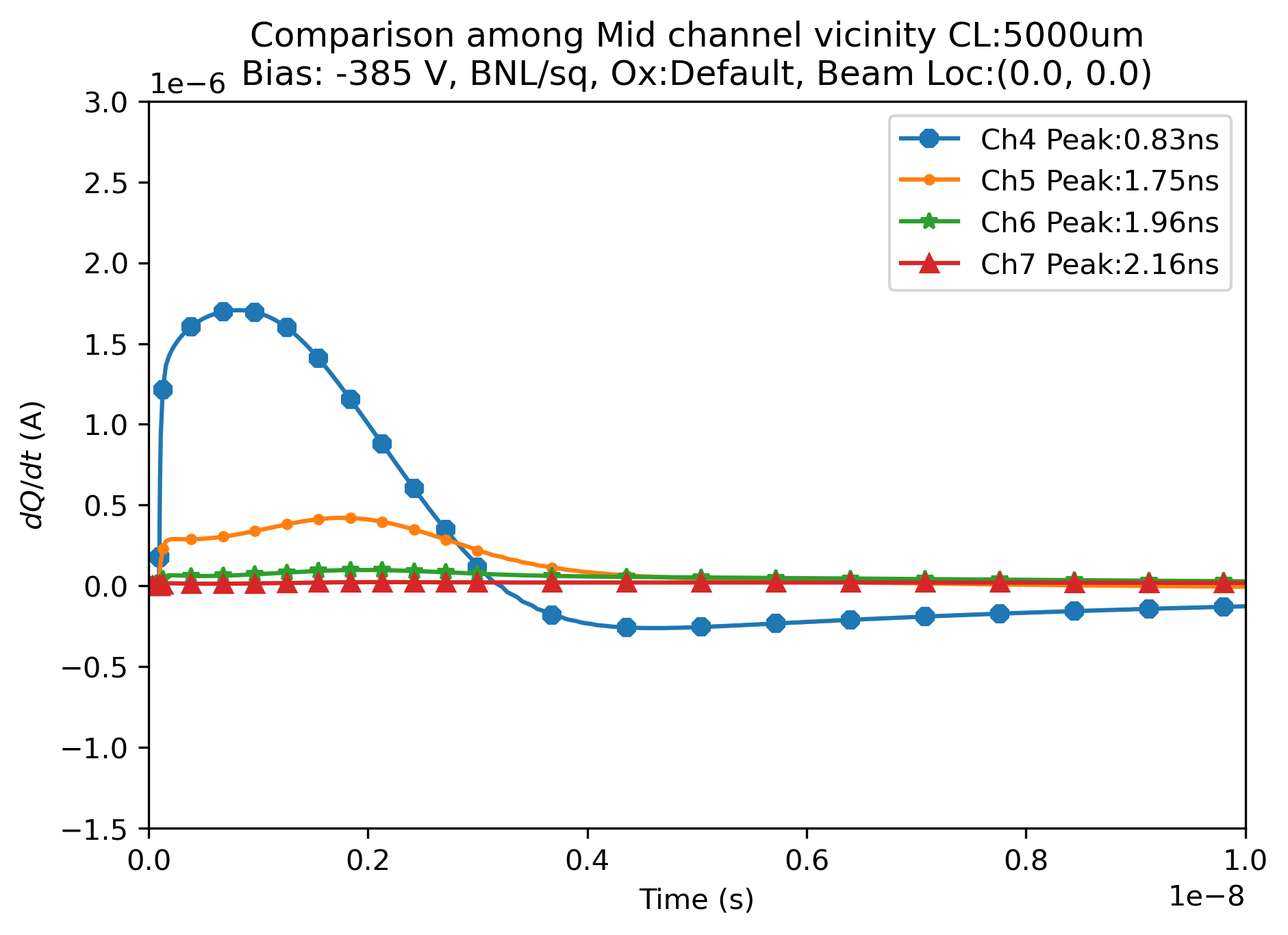}\label{fig:TCAD_Transients_120um_L5000}}
    \hfill
    \subfloat[L = 1 cm]{
        \includegraphics[width=0.45\textwidth]{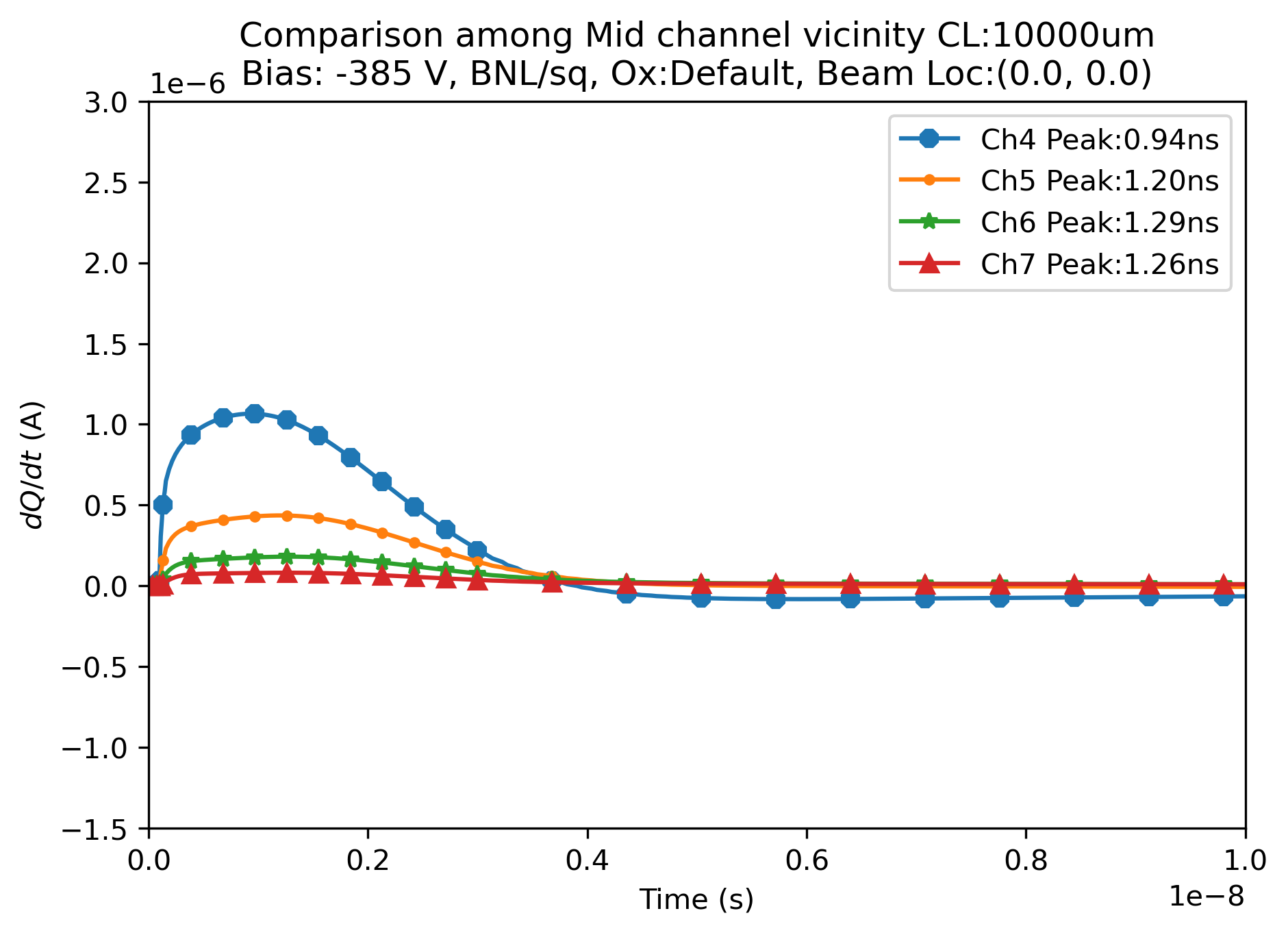}\label{fig:TCAD_Transients_120um_L1cm}}
   
    \caption{Transient responses of varying channel lengths of 120-\um-thick devices.}
    \label{fig:TCAD_Transients_120um}
\end{figure}

\paragraph{Bulk thickness}
Additionally, we have performed the beam scan simulation on a thicker substrate, the thickness of 120~\um, device, and their result is displayed in Fig.~\ref{fig:TCAD_Transients_120um}. Due to the thicker substrate, the breakdown point was around -400~V of back contact bias which was roughly 100~V lower than the 50~\um\ case. Thus, we decided to run the transient simulation at the bias point of -385~V. The tendency of decreasing peak current of channel 4 still persists in the 120-\um-thick cases. However, in the thicker devices, the first neighboring strip shows delayed responses, peaking at 2.42~ns in Fig.~\ref{fig:TCAD_Transients_120um_L500}, and such delayed neighboring peaks appear when the channel length of the device is shorter than 1~cm. At 1~cm, it seems the peak response delay of neighboring channels seems to disappear, unlike with the thinner, 50-\um-thick, device. In fact, the second neighbor of the 50~\um\ substrate shows similar behavior as seen in Fig.~\ref{fig:TCAD_Transients} where the influence of the middle channel is lower than the 120-\um-thick case. 

\begin{figure}[ht]
\centering
\includegraphics[width=0.85\textwidth]{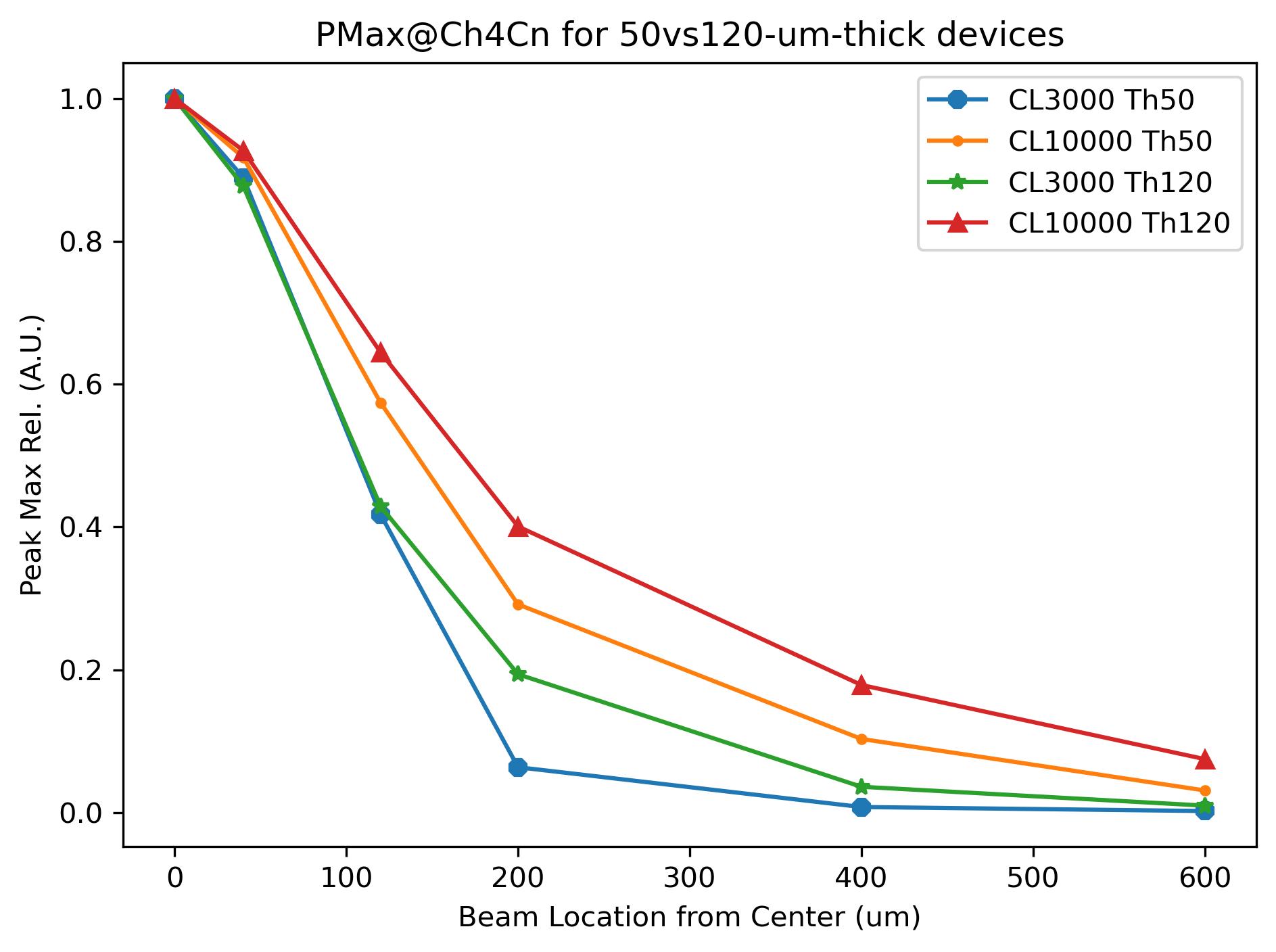}
\caption{P-Max plot of various strip lengths.}
\label{fig:TCAD_PMAX_thickness_comparison}
\end{figure}

Fig.~\ref{fig:TCAD_PMAX_thickness_comparison} shows the P-Max comparison between the 50-\um-thick substrate device and 120-\um-thick devices. The comparison was taken at channel lengths of 3000~\um\ and 1~cm. Here, 200~\um\ beam location is the first neighbor, 400~\um\ is the second neighbor, and 600~\um\ is the third neighbor, channel 7. Regardless of channel length, the thicker device shows more neighboring channel influence. However, the heightened influence due to the thickened substrate disappears as the distance from the mid-channel. At the first neighbor, at a shorter channel length of 3000~\um, the thicker device, 120-\um-thick, shows almost doubled crosstalk than the thinner device, yet, at the second and the third neighbors, the crosstalk decreases and even converges at the third neighbor. When the channel lengths are longer, the mid-channel influence is higher in the thicker device, yet the delta against the thinner substrate narrows, yet not as close to the shorter device. 

However, although the mid-channel influence converges down to minimal at further channels, the thickened substrate of AC-LGAD is detrimental to the spatial resolution of AC-LGADs, as seen in the Fig.~\ref{fig:TCAD_PMAX_thickness_comparison}. This can be due to the prolonged drift path of excited carriers in the substrate while the drift velocity of the carriers in a silicon substrate is limited. Since the diffusion of a carrier cloud is inevitable during the drift process, we can expect that the expanded carrier cloud due to diffusion may have caused the increased influences on neighbors. On the other hand, the increase of P-Max influence at the first neighbors seems to be limited when the device length is longer. At 3000-\um-long devices, the P-Max jumped almost twice while at the 1-cm-long device, the increase was around 23 \% at most.

\begin{figure}[ht]
\centering
\includegraphics[width=0.85\textwidth]{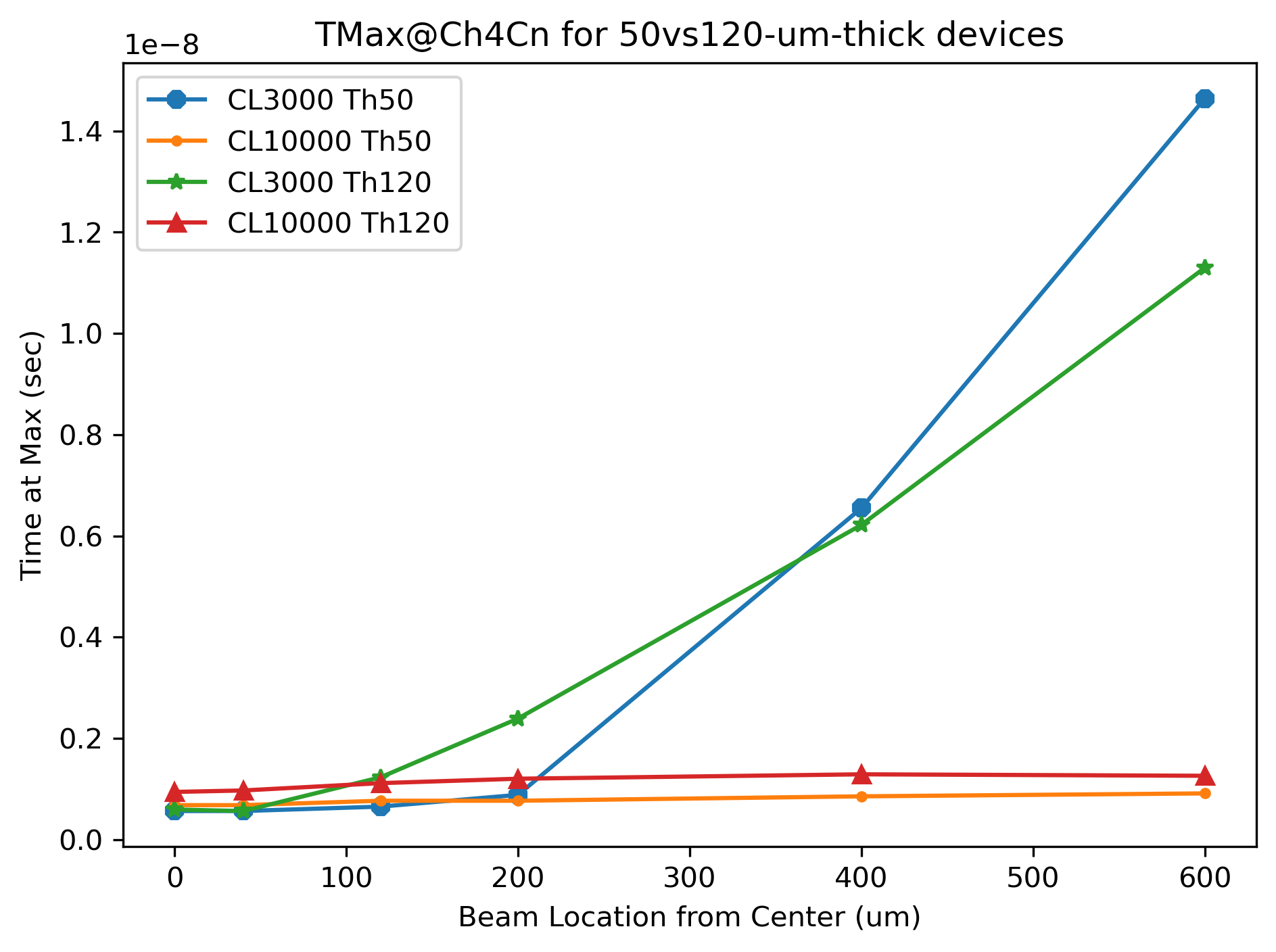}
\caption{T-Max plot of various strip lengths.}
\label{fig:TCAD_TMAX_thickness_comparison}
\end{figure}

Another aspect of the AC-LGAD readout is the T-Max, Time-at-Max, of a channel current. Fig.~\ref{fig:TCAD_TMAX_thickness_comparison} shows the T-Max comparison from the short/long and thick/thin devices we have discussed above. At shorter channel lengths, the time-at-max, T-Max increases as we get further from the middle channel, the influence of the channel appears slower at the remote channels. However, as the channel becomes longer, the overall device size is obviously larger than the shorter channel device, and the influence appears almost in sync. Note that the device's lateral aspect ratio at the 1000-\um-long device is 0.167. It can be assumed that the longer device may have been overwhelmed by carrier conduction parallel to the channels. But the carrier collection electrodes align parallel to the channels which means the carriers will be collected only by drifting to those electrodes, directions of perpendicular to the channel, rather than drifting along the channels. This behavior may need a more thorough investigation with carrier density snapshot at each transient point. 

\FloatBarrier
\section{Conclusions}
AC-LGAD simulations with TCAD Silvaco and TCAD Sentaurus were presented in the form of a charge-sharing profile for varying parameters. 
The following observations can be made.
The effect of the N+ resistivity was shown with TCAD Silvaco simulation; higher resistivity decreases the width of the charge-sharing profile.
The effect of strip length and bulk thickness was probed With TCAD Sentaurus. A longer strip increases the width of the charge-sharing profile, especially for strips longer than a cm.
The bulk thickness also influences the charge-sharing profile; a thicker bulk increases the width of the charge-sharing profile. 
The time-at-maximum (T-Max) as a function of distance behaves differently with the channel length, likely an effect tied to the large inter-strip capacitance of the longer strips. This behavior will need a more thorough investigation with carrier density snapshot at each transient point. 

TCAD simulation is an invaluable tool for understanding the effect of each sensor parameter on the charge-sharing profile and the general behavior of AC-LGADs. 
Future sensor productions will significantly benefit from the simulation effort regarding expected sensor performance and yield.

\section{Acknowledgments}
{
\small
This work was supported by the United States Department of Energy grant DE-FG02-04ER41286.

\bibliography{bib/TechnicalProposal,bib/hpk_fbk_paper,bib/HGTD_TDR,bib/SHIN,bib/nizam}
}

\end{document}